\documentclass{emulateapj}
\usepackage{apjfonts}

\newcommand{\etal}{et al.~}
\def\gsim{\lower 2pt \hbox{$\, \buildrel {\scriptstyle >}\over
{\scriptstyle \sim}\,$}}
\def\lsim{\lower 2pt \hbox{$\, \buildrel {\scriptstyle <}\over
{\scriptstyle \sim}\,$}}

\def\oviii{O~{\scriptsize VIII}}
\def\ovii{O~{\scriptsize VII}}
\def\ovi{O~{\scriptsize VI}}

\def\neviii{Ne~{\scriptsize VIII}}
\def\neix{Ne~{\scriptsize IX}}
\def\nex{Ne~{\scriptsize X}}

\shortauthors{Yao \etal}
\shorttitle{Limits on Hot Galactic Halo Gas}

\begin{document}

\slugcomment{\em Accepted for publication in the ApJ Letters}
\title{Limits on Hot Galactic Halo Gas from X-ray Absorption Lines}
\author{Y. Yao\altaffilmark{1}, 
        M. A. Nowak\altaffilmark{1},
	Q. D. Wang\altaffilmark{2},
	N. S. Schulz\altaffilmark{1},
	and
        C. R. Canizares\altaffilmark{1}}
\altaffiltext{1}{Massachusetts Institute of Technology (MIT) Kavli Institute
 for Astrophysics and Space Research, 70 Vassar Street, Cambridge, MA 02139;
 yaoys, mnowak, nss, and crc@space.mit.edu}
\altaffiltext{2}{Department of Astronomy, University of Massachusetts, 
  Amherst, MA 01003; wqd@astro.umass.edu}

\begin{abstract}

Although the existence of large-scale hot gaseous halos around massive disk
galaxies have been theorized for a long time, there is yet very little
observational evidence. We report the {\sl Chandra} and {\sl XMM-Newton} 
grating spectral detection of \ovii\ and \neix\ K$\alpha$
absorption lines along the sight-line of 4U~1957+11.
The line absorption is consistent with the interstellar medium in 
origin. Attributing these line absorptions to the hot gas associated with
the Galactic disk, we search for the gaseous halo around the Milky Way 
by comparing this sight-line with more distant ones (toward LMC X--3 
and the AGN Mrk~421).
We find that all the line
absorptions along the LMC X-3 and Mrk~421 sight-lines are attributable to
the hot gas in a thick Galactic disk, as traced by the absorption 
lines in the spectra of 4U~1957+11 after a Galactic latitude dependent
correction. We constrain the \ovii\ column density
through the halo to be $N_{\rm OVII}<5\times10^{15}~{\rm cm^{-2}}$
(95\% confidence limit), and conclude that the hot gas
contribution to the metal line absorptions, if existing, is negligible.

\end{abstract}


\section{Introduction }
\label{sec:intro}

Many semi-analytic calculations and numerical simulations 
for disk galaxy formation predict the
existence of extended hot gaseous halos around massive spirals due to
the accretion of the intergalactic medium. 
For the Milky Way, the gas temperature can be 
shock-heated to $\sim10^6$ K at the virial radius ($\sim250$ kpc;
e.g., Birnboim \& Dekel 2003; Fukugita \& Peebles 2006), and the total mass
contained in the large scale halo can be comparable with or
even greater than the total baryonic mass of stars and the interstellar medium
in the Galactic disk (e.g., Sommer-Larsen 2006). Clearly, an observational 
measurement of such an extended halo will provide an important test of
galaxy formation theories.

Searching for X-ray emission from large scale halos around nearby disk
galaxies has proven unsuccessful. Extraplanar X-ray emissions
have indeed been routinely detected around a number of nearby spirals, but only
on scales of several kpc (except for the star burst galaxies), and 
most likely as the result from on-going stellar feedback in galactic disks
(e.g., T\"ullmann \etal 2006; Li \etal 2006). 
The X-ray surface brightness of a galactic halo must be very weak at 
large scales, at least partly because of the density square dependence of 
the emission and the possible low metallicity of the gas. 
So far, the only claimed detection of 
an apparent X-ray-emitting halo on a scale of $\sim20$ kpc is around the 
quiescent edge-on disk galaxy NGC~5746 (Pedersen \etal 2006).

The large-scale hot gaseous halo around the Milky Way is indirectly 
evidenced by the presence of high velocity clouds (HVCs), in particular,
the detection of their associated \ovi\ line absorption 
(e.g., Spitzer 1956;  Sembach \etal 2003). 
Because the \ovi-bearing gas is preferentially populated at  
intermediate temperatures of $\sim 3 \times 10^5$ K where thermally 
unstable cooling occurs, 
it is believed to be produced at the interfaces between 
the cold/warm and the hot media. Without knowing the temperature
distribution and the metallicity of the hot gas, however, it is very hard 
to reliably estimate its total mass.

The best way to directly measure the hot medium is through its X-ray 
absorption line features. Indeed, \ovii, \oviii, and/or
\neix\ absorption lines consistent with zero velocity shifts ($cz\sim0$) 
have been detected unambiguously toward several bright extragalactic sources 
(e.g., LMC~X--3, Mrk~421, and PKS~2155-304; Nicastro \etal 2002; 
Fang \etal 2003; Wang \etal 2005; Yao \& Wang 2007a, YW07a hereafter). 
But these sight-lines pass through the Galactic disk, the extended 
Galactic halo, and (for those AGNs) the possible warm hot intergalactic 
medium (WHIM) in the Local Group. Recently, it has been argued that the bulk
of these absorptions can {\sl not} be produced in the WHIM 
\citep{fang06, yao07a, bre07}. Indeed, these highly ionized absorption lines 
have also been detected in spectra of many
Galactic sources 
(e.g., Yao \& Wang 2005; Juett \etal 2006), 
clearly indicating the existence of the hot gas in the Galactic disk.
All existing X-ray absorption and emission data 
are consistent with the hot gas located 
in the Galactic disk with a vertical exponential scale height of $\sim 2$ 
kpc (Yao \& Wang 2005, 2007a), comparable to
those inferred from the angular distributions of column densities 
of the \ovi-bearing gas and pulsar dispersion measures 
(Savage \etal 2003; Berkhuijsen \etal 2006). Clearly, the constraint on the
location of the hot gas depends on the modeling. For instance, 
\citet{bre07} found that the absorbing gas toward AGNs are
also consistent with a uniformly spherical distribution within a radius of 20 
kpc. The questions here are, whether the extended Galactic halo 
(defined as the hot gas in the region of $>$10 kpc beyond the Galactic plane
but within the Galactic virial radius)
exists, and how much does it contribute to the observed 
highly ionized X-ray absorptions?

In this Letter, we present a differential study searching for such a
large-scale hot gaseous halo around our Galaxy by comparing observations
between sight-lines of 4U~1957+11 and Mrk~421, and between those of LMC~X--3 
and Mrk~421. \neix, \ovii, and/or \oviii\ absorption lines at $cz\sim0$ 
have been detected in spectra of all three sources observed with 
{\sl Chandra} and/or {\sl XMM-Newton X-ray Observatories}. These comparisons  
enable us to probe the Galactic halo by examining its contribution to the 
observed absorption lines, and then 
to estimate its total mass and/or its metallicity.

Throughout the Letter, we adopt the solar abundances from \citet{wil00}, 
quote statistical errors (or upper limits) for single floating parameters at
90\% (95\%) confidence levels, and assume that the hot gas in both the Galactic
disk and halo is in the collisional ionization equilibrium state and is
approximately isothermal. We further assume that the disk gas is distributed 
following an exponential decay law with a vertical scale height of 2 kpc 
(e.g., Savage, \etal 2003; YW07a). All the data analyses are performed with 
the software package XSPEC (version 11.3.2).

\section{Observations and data reduction}
\label{sec:obs}

4U~1957+11 (V1408 Aquilae; $l, b=51\fdg31, -9\fdg33$) is a persistent low mass 
X-ray binary (Nowak \& Wilms 1999 and references therein). {\sl Chandra} 
observed this source for 67 ks on 2004 September 7 (ObsID 4552) with the 
High Energy Transmission Grating Spectrometer (HETGS; Canizares \etal 2005)
and {\sl XMM-Newton} observed it for 45 ks on 2004 October 16
(ObsID 206320101).

We reprocessed both {\sl Chandra} and {\sl XMM-Newton} observations following the 
standard procedures.
Using the software package {\sl CIAO} 
\footnote{http://cxc.harvard.edu/ciao/}
and calibration database (CALDB) version 3.4,
we re-calibrated the {\sl Chandra} observation, extracted grating spectra, and
calculated corresponding instrumental response files (RSPs). 
In this study, we only use the first order 
spectra from the medium energy grating because of its large effective 
area at longer wavelengths ($>13$ \AA). To further enhance the counting 
statistics, we co-added the positive- and the negative-grating spectra and 
RSPs. For the {\sl XMM-Newton} observation, we used the software package {\sl SAS} 
(version 6.50)
\footnote{http://xmm.vilspa.esa.es/sas/current/documentation/threads}
to remove those events contaminated with background flares
(by screening out the time intervals with an event count rate $>0.4$ counts/s 
on the CCD9), and then extracted
spectra and calculated RSPs from the Reflection Grating 
Spectrometer (RGS) events by running the thread {\sl rgsproc}.
Because of the failure of the CCD4 in the RGS2, we only use the RGS1 spectrum.
 
In order to measure absorption line properties we only rely on the nearby 
continuum levels. We therefore use several parts of the spectra that are local 
to the lines of our interest. 
For the {\sl Chandra} spectrum, we use a wavelength range of
12.0--22.5 \AA\ covering the K$\alpha$ and/or K$\beta$ transition lines of 
\ovii, \oviii\, \neviii, \neix, and \nex. 
For the case of the {\sl XMM-Newton} spectrum, 
the RGS1 has no effective area around the \neix\ K$\alpha$ line
and there is a bad pixel near the \oviii\ K$\alpha$ line
at 18.969 \AA. Here we use ranges of 18.0--18.9 and
19.2--22.5 \AA\ covering the \ovii\ K$\alpha$ and K$\beta$ lines.


Mrk~421, a bright quasar at $z=0.03$ ($l, b=179\fdg83, 65\fdg03$),
is a {\sl Chandra} calibration target and has been observed with the
high resolution grating instruments multiple times; most of the 
observations have been reported in YW07. 
In this Letter, we use the same observations, and the same spectra and 
RSPs as in YW07. To fit the 
spectra without worrying about overlapping spectral orders, for each 
of the observations taken with High Resolution Camera (HRC; Table~1 in YW07),
we derive the first order spectrum with minimal higher ($>1$) order 
confusion via the following steps. We first fit the broadband 
(1.5-35 \AA) order-overlapped spectrum with the order combined RSP, and then
multiply the spectral counts channel by channel with the ratio between the 
model predicated counts based on the first order RSP and that based on the
order-combined RSP. We then combine all the first order spectra of different
observations using the same procedure as in YW07.

In YW07, we have reported the \ovii\ K$\alpha$, K$\beta$, and \oviii\ K$\alpha$
absorption lines in the spectrum of Mrk~421 and measured their equivalent
widths (EWs; see also Williams \etal 2005; Kaastra \etal 2006). 
In fact, the \neix\ 
K$\alpha$ line has also been detected in the spectrum \citep{wil05}. However, 
we find that, due to an instrumental feature near the line, measuring 
the amount of the \neix\ absorption is severely affected by how the continuum 
is placed. To avoid such an uncertainty, in our analysis we only use the 
spectral range of 18--22.5 \AA, covering the oxygen lines as mentioned above.

LMC~X--3 ($l, b=273\fdg58, -32\fdg08$) is located $\sim50$ kpc away from the 
Sun and {\sl Chandra} observed it with the Low Energy Transmission Grating 
Spectrometer plus HRC for $\sim100$ ks in total. The interstellar \ovii\ and 
\neix\ K$\alpha$ absorption
lines observed in this sight-line have been reported by \citet{wang05}.
In this Letter, we use the same spectra and RSPs as in Wang \etal. 
Following the same procedure as for Mrk~421 spectra (see above), we obtained 
the co-added 
first order spectrum and the RSP. We use the spectral range of 12.0--22.5 \AA\
in the following analysis.

\section{Analysis and results}
\label{sec:results}

\begin{figure}
   \plotone{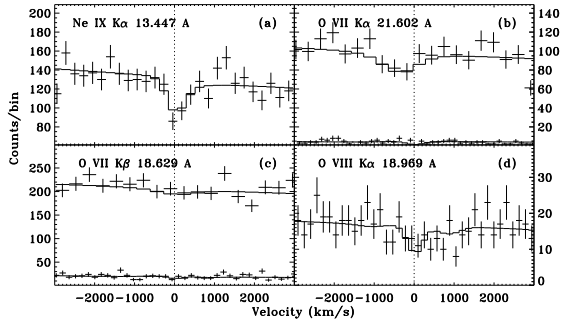}
   \caption{The oxygen and neon absorption lines detected in the spectra of 
   4U~1957+11. The upper plots in panels (b) and (c) are from the {\sl XMM-Newton}
   spectrum; the others are from the {\sl Chandra} spectrum. The vertical dotted 
   lines mark zero velocity.
   The bin-size is 22.2 m\AA\ and 10.0 m\AA\ for {\sl XMM-Newton} and
   {\sl Chandra} spectra, respectively.
    \label{fig:1957}
  }
\end{figure}

We focus our efforts on searching for and analyzing the highly 
ionized absorption lines at $cz\sim0$ that are likely produced in the hot 
ISM (\S~\ref{sec:dis}).
The \neix\ (13.447 \AA) and \ovii\ K$\alpha$ (21.602 \AA) lines consistent 
with $cz\sim0$ appear in the {\sl Chandra} and {\sl XMM-Newton} spectra
at $\sim3$ and $4\sigma$ significance levels (Fig.~\ref{fig:1957}), 
respectively. Modeling these two lines with negative
Gaussian models, 
we measure their EWs as 6.3(3.9, 8.8) and 
18.7(8.3, 30.6) m\AA.
Note that the EWs measured here are for reference only. The ionic column 
density and the dispersion velocity of the absorbing gas will be measured
in below. No other highly ionized O and Ne lines are
detected at $\gsim2\sigma$ significance. Fixing the line centroid at the 
rest frame wavelength \citep{ver96, beh02}, we obtain the EW upper limits 
of \nex (12.134 \AA), \neviii\ (13.646 \AA) and 
\oviii\ K$\alpha$ (18.969 \AA), and \ovii\ K$\beta$ (18.629 \AA) lines 
as 1.2, 2.8, 15.6, 10.1 m\AA, respectively.
Since all these line are unresolved, we fix the line width
$\sigma$ to 100 km~s$^{-1}$ (please refer to $v_b$ in Table~\ref{tab:results})
during these measurements and we use either the {\sl Chandra} or {\sl XMM-Newton} 
spectrum, whichever has higher counting statistics
near the line (Fig.~\ref{fig:1957}). 

A comparison of the line absorption between 4U~1957+11 and Mrk~421 sight-lines
provides an effective probe of the large scale Galactic hot gaseous halo.
The distance of 4U~1957+11 is estimated to be $D\sim10-25$ kpc 
(M. Nowak \etal in preparation), a location of 1.6 -- 4.1 kpc below
the Galactic disk plane where it samples 55-87\% of the Galactic 
disk gas in the vertical direction (\S~\ref{sec:intro}) 
along the line of sight. In comparison, the 
line of sight towards Mrk~421 samples the hot gas not only in the Galactic disk but 
also in the putative Galactic halo (\S~\ref{sec:intro}). Therefore the 
differential absorptions between these two sight-lines allow us to 
directly constrain the absorption contribution from the halo.

We first characterize the hot absorbing gas towards these two sight-lines by 
using our absorption line model, {\sl absline}. This model adopts the the 
{\sl Voigt} function to approximate an individual line profile, and uses the 
line centroid $E_l$, velocity dispersion $v_b$, absorbing gas temperature 
$T$, and reference ionic column density $N{\rm_X}$ (e.g., $X$=\ovii\ or H) 
as fitting parameters. It therefore can be used to conduct a joint analysis 
of multiple absorption lines (including non-detections).
\footnote{Please refer to \citet{yao05,yao06} for the model and 
the joint analysis procedure.} For the sight-line toward Mrk~421, 
we use \ovii\ K$\alpha$, K$\beta$, and \oviii\ K$\alpha$ absorption lines 
from the {\sl Chandra} spectrum (YW07). For sight-line toward 4U~1957+11, 
we use \ovii, \oviii, \neviii, \neix, \nex\ K$\alpha$, and \ovii\ K$\beta$ 
lines in the {\sl Chandra} and {\sl XMM-Newton} spectra where 
applicable, and assume the neon to oxygen abundance ratio  
to be the solar value \citep{yao06}.
From the joint analysis, we obtain the $T$, $v_b$, and $N_{\rm OVII}$ 
(or its equivalent $N_{\rm H}$ for a given 
metallicity), as reported in Table~\ref{tab:results}.  For ease of comparison, 
we take into account the dependence of the column density on the Galactic 
latitude (sin$b$ factor; \S~\ref{sec:intro}) for the 4U~1957+11 sight-line 
with respect to the Mrk~421 direction (Table~\ref{tab:results}).

\begin{deluxetable}{lcccc}
\tablewidth{0pt}
\tablecaption{
  Line Analysis Results\label{tab:results} }
\tablehead{
 & T          & $v_b$         & N$_{\rm OVII}$       & N$_{\rm H}$A$_{\rm O}$\\
 &($10^6$ K)  &(km s$^{-1}$)  &($10^{15}$ cm$^{-2}$) & ($10^{19}$ cm$^{-2}$) }
\startdata
Mrk~421$^a$  & 1.4(1.3, 1.6) & 64(48, 104)   & 10.0(6.6, 14.7)  &  1.4(1.0, 2.0) \\
4U~1957$^a$  & 2.2(1.6, 2.7) & 155(70, 301)  & 3.1(1.8, 8.2)  &  1.1(0.6, 1.8)\\
LMC~X3$^a$   & 1.3(0.8, 2.0) & 79(62, 132)   & 11.7(7.4, 18.5) & 2.3(1.5, 3.7) \\
\hline
4U~1957$^b$  & 1.8(1.7, 2.1) & 70(50, 172)   & 7.0(3.2, 12.7)  &  1.4(0.7, 2.2)\\
Halo$^c$ & $\cdots$      & $\cdots$      & $<4.8$         & $<0.9$ \\
\hline
LMC~X3$^b$ & 1.8(1.6, 2.0) & 64(49, 108)  & 8.8(5.8, 12.7) & 1.7(1.2, 2.2) \\
Halo$^c$ & $\cdots$      & $\cdots$      & $<3.7$         & $<0.7$ 
\enddata
\tablecomments{The uncertainty ranges (upper limits) are quoted at 
90\% (95\%) confidence levels. 
The dependence of the column density on the Galactic latitude
(sin$b$ factor) has been corrected with respect to Mrk~421 direction. 
$A_{\rm O}$ is the oxygen abundance in unit of the solar value. 
$^a$ Results from fitting lines in each direction separately.
$^b$ Results from a joint analysis of lines in the source and those in Mrk~421.
$^c$ The absorption contribution from the halo gas beyond the source.
See text for detail.
}
\end{deluxetable}

Please note that in the above characterization we assume 
that the intervening 
hot gas in both sight-lines is isothermal. In reality, the temperature of both
the disk and the halo gas could be a function of the vertical off-plane 
distance, the radial off-Galactic center distance, or a combination
of both (e.g., Toft \etal 2002; YW07).
This simplified description aims to provide
a direct comparison of the amount of absorption between the two sight-lines. 

\begin{figure}
   \plotone{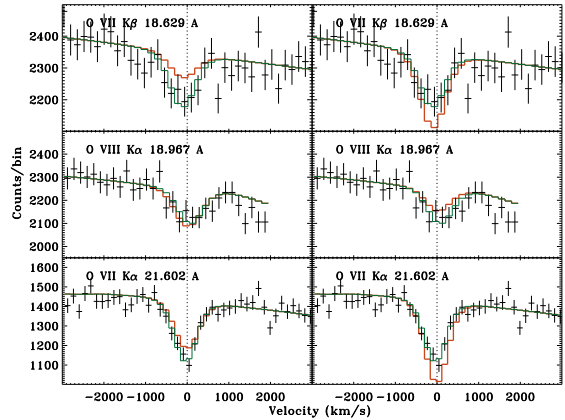}
   \caption{The observed Mrk~421 spectrum ({\sl cross}) around the oxygen 
   absorption lines and the best fit model ({\sl green histogram}) convolved
   with the instrumental response. The red histograms mark the amount of 
   absorption predicted for the Mrk~421 direction from the best 
   fit of the lines observed in the spectra of 4U~1957+11 ({\sl left panels})
   and LMC~X--3 ({\sl right panels}).
    \label{fig:mrk}
  }
\end{figure}

As demonstrated in Table~\ref{tab:results} and Figure~\ref{fig:mrk}, the 
absorption towards 4U~1957+11 can account for the bulk
of the absorption towards Mrk~421. To quantify the absorption
contribution from the putative Galactic halo gas, we
jointly analyze the absorption lines in these two directions,
``subtract'' the absorption spectrum toward 4U~1957+11 direction from that 
toward Mrk~421 direction, and attribute the residual absorption to 
the halo gas. We find that the halo gas is consistent with {\it no}
contribution to the observed line absorptions. Assuming that the halo
gas has the same thermal (e.g., $T$) and dynamic (e.g., $v_b$) properties 
as the disk gas, we estimate the column density upper limit of the halo gas 
as $N_{\rm OVII}<4.8\times10^{15}\ {\rm cm^{-2}}$ 
(or $N_{\rm H}A_{\rm O} < 9\times10^{18}\ {\rm cm^{-2}}$, where $A_{\rm O}$ is
the oxygen abundance in unit of the solar value; Table~\ref{tab:results}).
The halo gas and the disk gas may have different $T$; for the halo gas over
a broad temperature range of $0.4-2.5\times10^6$ K,
we find $N_HA_O<10^{19}~{\rm cm^{-2}}$. 

Following the same procedure as above, 
we characterize the line absorptions toward LMC~X--3 (Table~\ref{tab:results}; 
see also Wang \etal 2005), and then search for the absorption feature of the 
halo gas beyond LMC~X--3 by comparing it with Mrk~421. Again, we find that 
the line absorptions toward LMC~X--3 can account for all the absorptions 
observed toward the Mrk~421 sight-line (Fig.~\ref{fig:mrk}),
and obtain the upper limits of the column density of the halo by jointly 
analyzing the sight-lines of LMC~X--3 and Mrk~421 
(Table~\ref{tab:results}).

\section{Discussion}
\label{sec:dis}

In this study, we have assumed that the absorption lines observed 
in the spectra of 4U~1957+11 are produced in the ISM. However, the 
photo-ionized in- or out-flow material intrinsic to 4U~1957+11 could 
contaminate these lines.
From the best fit to the {\sl Chandra} spectrum in the range of 
2--23 \AA, 
we estimate the source luminosity over 0.5--10 keV as
$L=1.19D_{\rm 10kpc}^2\times10^{37}~{\rm ergs~s^{-1}}$, where $D_{\rm 10kpc}$
is the source distance in units of 10 kpc.
A curve of growth analysis with a dispersion velocity ($v_b$) of 
0 km~s$^{-1}$
gives a firm upper limit of the \neix\ column density to be
$N_{\rm NeIX}<6.6\times10^{17}~{\rm cm^{-2}}$ (assuming the solar abundance 
of Ne). This is equivalent to a
hydrogen column density $N_{\rm H}<1.1\times10^{22}~{\rm cm^{-2}}$,
assuming an ionization fraction of 0.5 for \neix. For the absorber,
the ionization parameter $U=L_x/n(r)r^2$ must satisfy $\log(U)\leq2.5$, since
the \nex\ K$\alpha$ absorption line has not been detected~\citep{kal82}.
Further assuming a radial gas density distribution
$n(r)\propto(r_w/r)^2$, where $r_w$ is the location where wind launches
and $r>r_w$,
we obtain $r_w > 50 R_\odot$, which is much larger than
the star separation ($\lsim6 R_\odot$) of the system
(estimated from its orbital period 9.33 hours and taking an
upper limit of the X-ray compact object $M_X<16 M_\odot$). These estimates
make the in-flow scenario very unlikely. On the other hand, if the absorber is
out-flow material, the lines are expected to be blue-shifted at velocities
larger than the escape velocity of the system (e.g., GRO J1655--40;
Miller \etal 2006), which is 260 ${\rm km\ s^{-1}}$ for 4U~1957+11
(taking $M_X=10 M_\odot$ and $r_w = 6 R_\odot$). In contrast, the detected
\neix\ K$\alpha$ velocity is $0\pm114~{\rm km~s^{-1}}$.
The above arguments suggest that the absorption lines are more likely 
produced in the ISM (see also Juett \etal 2006), 
although the intrinsic scenario can not be completely ruled out.

We have compared the highly ionized line absorptions of \ovii, 
\oviii, and \neix\ observed in the spectra of 4U~1957+11 and Mrk~421, and 
in the spectra of LMC~X--3 and Mrk~421 to search for the 
large scale Galactic halo. We found that there is no significant 
X-ray absorption due to hot gas beyond either 4U~1957+11 or LMC~X--3. 
These results are consistent with
our previous conclusion that bulk of the absorptions observed toward 
the Mrk~421 
sight-line are due to the hot gas around the Galactic disk with a scale height
of $\sim2$ kpc based on a joint analysis of the absorption and emission 
data (YW07). 
Note that the upper limits obtained toward LMC~X--3 
are slightly lower than those toward 4U~1957+11
(Table~\ref{tab:results}). Some of the difference are
due to the partial sampling of the Galactic disk by the latter sight-line, and
the better spectral quality of LMC~X--3 also matters.

Large scale hot gaseous halos are commonly expected in 
modern disk formation models for massive spirals (e.g. White \& Frenk 1991).
For the Milky Way in particular, the hot gaseous halo can extend up to 
$r_{\rm virial}\sim 250$~kpc and the total mass contained is 
expected to be $\sim6\times10^{10}\ M_\odot$, which is currently missing
in the baryon mass inventory in the Lambda Cold Dark Matter ($\Lambda$CDM)
cosmology
\citep{deh98, kly02, som06}. Recently, \citet{mal04} assumed that the 
fragmentated warm clouds could be formed out of the hot gas halo and
proposed a multiphase cooling scenario for the galaxy formation. This model
alleviates the so-called ``over-cooling'' problem faced by the 
standard cooling model (e.g., Klypin \etal 2002). They then obtained
a density profile for the residual hot halo gas as a function of radius
(Eq. 21 and Fig. 4 in Maller \& Bullock 2004). Normalizing their density 
profile by requiring the total column density integrated from 10 to 250 kpc
to be $N_{\rm H}A_{\rm O} < 10^{19}\ {\rm cm^{-2}}$ (Tabel~\ref{tab:results}),
we obtain the total mass contained in the halo as
$M_{\rm H}<1.2/A_{\rm O}\times10^{10}\ M_\odot$. Clearly, to match this upper
limit with the total missing baryon mass in the MW, 
$A_{\rm O}$ should be $<20\%$ 
of the solar value. More recently, \citet{han06} assumed that gaseous
halos of galaxies are 
in a hydrostatic equilibrium state and that the gas density and the total 
mass density profiles are power laws, and then derived a hot halo gas density 
profile for the MW-like galaxy (Fig.~2 in Hansen \& Sommer-Larsen 2006). 
Again, normalizing their profile we obtain the 
total hot gas mass as $<2.2/A_{\rm O}\times10^9$ M$_\odot$, requiring
$A_{\rm O}$ to be $<3.7\%$ of the solar value.

Low metallicity of the hot halo gas is not surprising; the halo gas is 
believed to be primarily accreted from the intergalactic medium that
is expected to be metal poor. This would also suggest that the 
star formation and the associated stellar feedback via Type II supernovae 
(SNeII) that occurred in the Galactic disk and bulge 
are not important in regulating the metal content of the halo.
Hot gas resulting from SNeII, if leaking into the halo 
(e.g., Mac Low \& Ferrara 1999), may cool quickly enough (via adiabatic 
expansion and atomic line radiation) to form metal-rich cold clouds in the 
local environment before homogenizing with
the halo gas (Wang 2007). As these clouds return back
to the disk plane as Galactic fountains, their interaction 
with the embedding hot gas could be another mechanism of warm cloud 
formation as 
required in multiphase cooling scenarios \citep{mal04}. The interfaces 
between these two media may harbor the bulk of HVCs as observed via \ovi\ 
absorptions at large distance (e.g., Sembach \etal 2003).

In this Letter, we derive the mass and metallicity limits on the putative 
Galactic halo by comparing the observed absorption lines among the sight-lines
of 4U~1957+11, LMC~X--3, and Mrk~421. However, the amount of absorptions 
observed along different AGN sight-lines could vary substantially 
(e.g., Bregman \& Lloyd-Davies 2007), which may be due to additional 
absorption components (e.g., 3C~273; Yao \& Wang 2007b) and/or different
absorbing gas properties. More comparisons among other sight-lines are 
therefore needed to confirm the presented results.

\acknowledgements 

We are grateful to Li Ji and the anonymous referee for their insightful 
comments, which helped to improve the presentation of the paper.
This work is supported by NASA through the Smithsonian Astrophysical 
Observatory contract SV3-73016 to 
MIT for support of the Chandra X-Ray Center 
under contract NAS 08-03060. 
Support from SAO/CXC 
grants AR6-7023 and AR7-8014 are also acknowledged.

\end{document}